\documentclass[12pt,epsf,amstex]{article}
\usepackage [dvips]{graphicx}
\usepackage{amsmath}
\usepackage{amssymb}
\usepackage{color}
\usepackage{epsfig}

\addtocounter{secnumdepth}{1}
\begin{document}
\newcommand{\br}{\bar{r}}
\newcommand{\bbeta}{\bar{\beta}}
\newcommand{\bgamma}{\bar{\gamma}}
\newcommand{\bE}{{\bf{E}}}
\newcommand{\bR}{{\bf{R}}}
\newcommand{\bS}{{\bf{S}}}
\newcommand{\bT}{\mbox{\bf T}}
\newcommand{\bt}{\mbox{\bf t}}
\newcommand{\bRn}{\bR^{[n]}}
\newcommand{\bRnbt}{\bR^{[n]}_{\mathbf t}}
\newcommand{\bRprimen}{\bR^{\prime\,[n]}}
\newcommand{\bRprimenmbt}{\bR^{\prime\,[n]}_{-\mathbf{t}}}
\newcommand{\bRav}{\bR_{\rm av}}
\newcommand{\half}{\frac{1}{2}}
\newcommand{\summ}{\sum_{m=1}^n}
\newcommand{\sumqno}{\sum_{q\neq 0}}
\newcommand{\tsum}{\Sigma}
\newcommand{\bsA}{\mathbf{A}}
\newcommand{\bsV}{\mathbf{V}}
\newcommand{\bsE}{\mathbf{E}}
\newcommand{\bsT}{\mathbf{T}}
\newcommand{\bsZ}{\hat{\mathbf{Z}}}
\newcommand{\bsPhi}{\mathbf{\Phi}}
\newcommand{\bse}{\mbox{\bf{1}}}
\newcommand{\bsphi}{{\boldsymbol{\varphi}}}
\newcommand{\bspsi}{{\boldsymbol{\psi}}}
\newcommand{\cdottt}{\!\cdot\!}
\newcommand{\deltaR}{\delta\mspace{-1.5mu}R}

\newcommand{\bGamma}{\boldmath$\Gamma$\unboldmath}
\newcommand{\dd}{\mbox{d}}
\newcommand{\ee}{\mbox{e}}
\newcommand{\p}{\partial}

\newcommand{\Rav}{R_{\rm av}}
\newcommand{\Rc}{R_{\rm c}}

\newcommand{\la}{\langle}
\newcommand{\ra}{\rangle}
\newcommand{\rao}{\rangle\raisebox{-.5ex}{$\!{}_0$}}  
\newcommand{\rae}{\rangle\raisebox{-.5ex}{$\!{}_1$}}
\newcommand{\raG}{\rangle_{_{\!G}}}
\newcommand{\rainr}{\rangle_r^{\rm in}}
\newcommand{\beq}{\begin{equation}}
\newcommand{\eeq}{\end{equation}}
\newcommand{\bea}{\begin{eqnarray}}
\newcommand{\eea}{\end{eqnarray}}
\def\lsim{\:\raisebox{-0.5ex}{$\stackrel{\textstyle<}{\sim}$}\:}
\def\gsim{\:\raisebox{-0.5ex}{$\stackrel{\textstyle>}{\sim}$}\:}

\numberwithin{equation}{section}

\thispagestyle{empty}
\title{\Large {\bf 
Random line tessellations of the plane:\\[3mm]
statistical properties of many-sided cells\\
\phantom{xxx} }}
 
\author{{H.\,J. Hilhorst$^1$  and P. Calka$^2$}\\[5mm]
{\small $^1$Laboratoire de Physique Th\'eorique, B\^atiment 210}\\[-1mm] 
{\small Univ Paris-Sud and CNRS,\, 91405 Orsay Cedex, France}\\
{\small $^2$Laboratoire MAP5, Universit\'e Paris Descartes, 45, rue des
  Saints-P\`eres}\\[-1mm] 
{\small 75270 Paris Cedex 06}\\}

\maketitle
%\vspace{-1cm}
\begin{small}
\begin{abstract}
\noindent 

We consider a family 
of random line tessellations of the Euclidean plane
introduced in a much more formal context
by Hug and Schneider [{\it Geom.~Funct.~Anal.} {\bf 17}, 156 (2007)]
and described by a parameter $\alpha\geq 1$.
For $\alpha=1$ the zero-cell 
(that is, the cell containing the origin) 
coincides with the Crofton cell of a Poisson line tessellation,
and for $\alpha=2$ it coincides with the typical Poisson-Voronoi cell.
Let ${p}_n(\alpha)$ 
be the probability for the zero-cell to have $n$ sides. 
By the methods of statistical mechanics
we construct the asymptotic expansion of 
$\log {p}_n(\alpha)$ up to terms that vanish as $n\to\infty$.
In the large-$n$ limit the cell is shown to become circular. 
The circle is centered at the origin when $\alpha>1$,
but gets delocalized for the Crofton cell, $\alpha=1$, 
which is a singular point of the parameter range.
The large-$n$ expansion of $\log p_n(1)$ 
is therefore different from that of the general case
and we show how to carry it out. 
As a corollary we obtain the analogous expansion for
the {\it typical\,} $n$-sided cell of a Poisson line tessellation.
\\

\end{abstract}
\end{small}
\vspace{15mm}

\noindent LPT Orsay 08-17\\
\newpage

%%%%%%%%%%%%%%%%%%%%%%%%%%%%%%%%%%%%%%%%%%%%%%%%%%%%%%%%%%%%%%%%%%%%%%%%%%%%%
%%%%%%%%%%%%%%%%%%%%%%%%%%%%%%%%%%%%%%%%%%%%%%%%%%%%%%%%%%%%%%%%%%%%%%%%%%%%%
%%%%%%%%%%%%%%%%%%%%%%%%%%%%%%%%%%%%%%%%%%%%%%%%%%%%%%%%%%%%%%%%%%%%%%%%%%%%%

\section{Introduction} 
\label{secintroduction}
\vspace{5mm}

%%%%%%%%%%%%%%%%%%%%%%%%%%%%%%%%%%%%%%%%%%%%%%%%%%%%%%%%%%%%%%%%%%%%%%%%%%%%%

\subsection{Poisson-Voronoi and Poisson line tessellations}
\label{sectessellations}

A {\it Voronoi tessellation\,} of ${\mathbb R}^2$  
is a partitioning of the plane into 
cells constructed around `point particles' 
in such a way that each point of space
is in the cell of the particle to which it is closest. 
When the point particle distribution is uniformly random,
(or, in mathematical terms, when it corresponds to 
a homogeneous {\it Poisson point process\,}),
the resulting partition is said to be a {\it Poisson-Voronoi tessellation}.
It is one of the simplest mathematical models of 
naturally occurring planar cellular structures.
Because of the great variety of their applications,
the statistics of Poisson-Voronoi cells has been 
studied in many different areas of science.
References may be found in 
the encyclopedic review by Okabe {\it et al.} \cite{Okabeetal00}.

\begin{figure}
\begin{center}
\scalebox{.45}
{\includegraphics{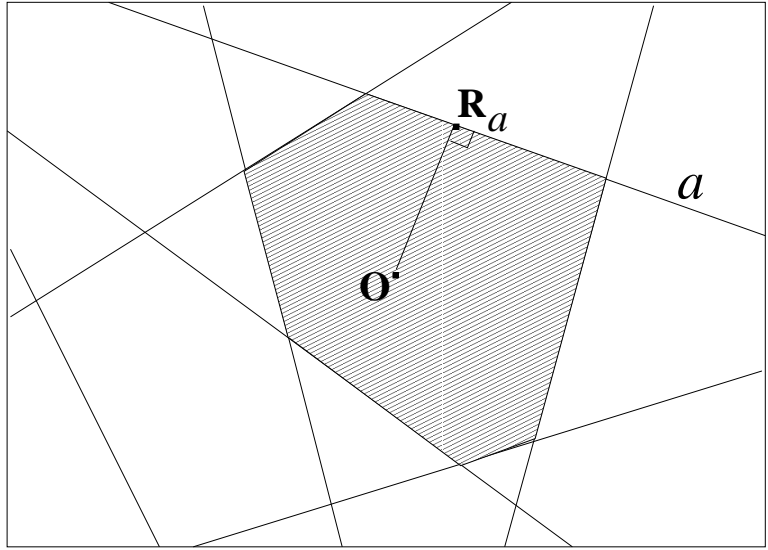}}
\end{center}
\caption{\small Example of a line tessellation of the plane. 
The projection of the origin onto line $a$
is denoted $\bR_a$; this projected point fully determines $a$. 
The cell surrounding the origin (the `zero-cell') has been shaded.}
\label{figlinetess}
\end{figure}

A different way of partitioning the plane into cells
is by means of intersecting straight lines, as in figure \ref{figlinetess}.
%%%%%%%%%%%%%%%%%%%%%%%%%%%%%%%%
%%%%%%%%%%%%%%%%%%%%%%%%%%%%%%%%
%%%%%%%%%%%%%%%%%%%%%%%%%%%%%%%%
%%%%%%%%%%%%%%%%%%%%%%%%%%%%%%%%
When these have a uniform distribution 
(or, in mathematical terms, when they correspond to a homogeneous 
{\it Poisson line process\,}), we refer to the partition
as a {\it Poisson line tessellation}.
The Poisson-Voronoi tessellation and the Poisson line tessellation
are both statistically invariant under translations and rotations in the
plane. For both, the cells are convex polygons.

An early application of the 
Poisson line tessellation occurs in work by Goudsmit \cite{Goudsmit45},
carried out at the suggestion of Niels Bohr. 
The question in that paper arose from cloud chamber experiments:
when three lines seemingly originate from the same point, then what is the
probability that they do not result from the same event?
Hence the problem became to calculate the probability for three
independent lines to nearly pass through the same point,
or, put differently, for a typical triangular cell to have an area less than
$A$ in the limit of very small $A$. 
Poisson line tessellations have since interested mathematicians with
important contributions due to, in particular,
Miles \cite{Miles64a,Miles64b,Miles73}, Matheron \cite{Matheron75}, Kovalenko
\cite{Kovalenko98,Kovalenko99}, Goldman \cite{Goldman98}, and Hug, Reitzner 
and Schneider \cite{Hugetal04}.  

The cell that contains the origin is generally called the `zero-cell'.
In the Poisson line tessellation it carries the special name 
of `Crofton cell', 
in reference to Crofton's formula in integral geometry \cite{footnoteCr}.
Since the origin falls in a cell of area $A$ with a probability 
proportional to $A$, the Crofton cell is not typical but
more likely larger-than-typical. 
In mathematical terms, the density of the typical cell 
differs from that of the Crofton cell by a factor $A/\la A \ra^{\rm typ}$,
where $\la A \ra^{\rm typ}$ is the average typical-cell area.
We will distinguish quantities pertaining to the typical cell of a Poisson
line tessellation (as opposed to the Crofton cell) by an extra superscript 
`typ'.

%%%%%%%%%%%%%%%%%%%%%%%%%%%%%%%%%%%%%%%%%%%%%%%%%%%%%%%%%%%%%%%%%%%%%%%%%%%

\subsection{Cell sidedness}
\label{secplanar}

The cell property most studied
is the sidedness probability $p_n$, that is, the probability
for the cell to have $n$ sides.
Other quantities of interest have included
the averages, moments, and correlations of $n$, the cell area and
the cell perimeter, as well as the 
distribution of the angles between the perimeter segments.
The statistical properties of an $n$-sided cell
may be expressed analytically 
as $2n$-fold integrals on the planar coordinates of the point particles
(for the Voronoi tessellation) or of the lines (for the line tessellation)
defining that cell. 
However, only few of these integrals can be evaluated exactly.
In particular the calculation of the fraction $p_n$
has so far been impossible for general $n$, whether for the 
Poisson-Voronoi or for the Poisson line tessellation.

We briefly recall some known results.
In a Poisson line tessellation the typical cell 
has an average number of sides $\la n \ra^{\rm typ}=4$.
The sidedness $p^{\rm typ}_n$ peaks at $n=4$. 
Miles \cite{Miles64a} obtained the exact value 
$p^{\rm typ}_3=2-\pi^2/6 = 0.35506...$ 
and Tanner \cite{Tanner83a} showed that 
$p^{\rm typ}_4 =
\pi^2\log 2 - \tfrac{1}{3} - \tfrac{7}{36}\pi^2 - \tfrac{7}{2}\zeta(3)
= 0.38146...$. 
Numerical values for $p^{\rm typ}_n$ based on Monte Carlo simulation
were given by Crain and Miles \cite{CrainMiles76},
by George \cite{George87}, 
and recently by Michel and Paroux \cite{MichelParoux07}
(who compare their results to the earlier ones),
for sidednesses not exceeding $n=12$.

For the Crofton cell, by contrast,
Matheron \cite{Matheron75} showed that 
the average sidedness is $\la n \ra = \pi^2/2 = 4.9348...$.
The distribution $p_n$ peaks at $n=5$.
Miles \cite{Miles73} obtained the only known exact result, namely 
$p_3 = (25-36\log 2){\pi^2}/6 = 0.076820...$.
Numerical values for $p_n$
were given by Calka \cite{Calka03a} for $n=3,4,\ldots,9$ and
by Michel and Paroux \cite{MichelParoux07} for $n=3,4,\ldots,11$.

The Poisson-Voronoi cell has an average sidedness of $\la n \ra = 6$.
Its sidedness distribution $p_n$ peaks at $n=6$.
Although no exact results are known for any of the $p_n$, 
it has recently been possible \cite{Hilhorst05a,Hilhorst05b,Hilhorst08}
to obtain the asymptotic expansion of $p_n$ 
in the limit of asymptotically large $n$. The expansion was shown to have
implications for the asymptotic cell shape, as well as for
the finite $n$ behavior \cite{Hilhorst07} 
and for correlations between neighboring cells \cite{Hilhorst06}.

A natural question to ask, then, is whether 
a similar asymptotic analysis of $p_n$ (for the Crofton cell) and
of $p^{\rm typ}_n$ (for the typical cell) 
can be carried out for the Poisson line tessellation. 
Now it was observed by Hug and Schneider \cite{HugSchneider07} that
the Crofton cell and the typical Poisson-Voronoi cell  
are particular instances of a more general family of
zero-cell problems, dependent on a parameter $\alpha$.
Hence our interest in the Poisson line tessellation leads us 
quite naturally to study here this full one-parameter family,
which we will define in the next subsection.

%%%%%%%%%%%%%%%%%%%%%%%%%%%%%%%%%%%%%%%%%%%%%%%%%%%%%%%%%%%%%%%%%%%%%%%%%%%%%
%%%%%%%%%%%%%%%%%%%%%%%%%%%%%%%%%%%%%%%%%%%%%%%%%%%%%%%%%%%%%%%%%%%%%%%%%%%%%

\subsection{A family of tessellations}
\label{secintrointerpol}

Let $a$ be a line in the plane and $\bR_a$ the projection of the origin
onto that line, as shown in figure \ref{figlinetess}; 
then $a$ is uniquely specified by $\bR_a$. 
In the Poisson line tessellation the $\bR_a$ 
are independent identically distributed stochastic vectors with 
a density proportional to $1/R_a$.
This tessellation is statistically
invariant under translations. In fact, the intersections of the 
lines $a$ of the tessellation with
an arbitrary additional line constitute a one-dimensional Poisson process;
and the associated angles of intersection $\theta_a$ 
are mutually independent and have the common
probability density $\frac{1}{2}\sin\theta$, where $0<\theta<\pi$. 
In this work we will consider
a more general tessellation that was introduced by Hug and Schneider
\cite{HugSchneider07} and depends on a parameter $\alpha$.
The projection vectors of this tessellation are distributed with a density
\beq
\rho(\bR) = \mbox{cst} \times R^{\alpha-2}, \qquad \alpha \geq 1. 
\label{defrho}
\eeq
For generic $\alpha$ the distribution (\ref{defrho}) has central 
symmetry around the origin
but the corresponding tessellation is not translationally invariant.
The zero-cell is therefore unlike any other cell.
For $\alpha=1$ we recover the Poisson line tessellation. 
For $\alpha=2$ the zero-cell corresponding to (\ref{defrho}) 
is identical to the {\it typical\,} cell of the Poisson-Voronoi
tessellation, which is easily seen as follows.
For $\alpha=2$ we have $\rho(\bR)=\mbox{cst}$,
so that the projections themselves have a uniform density.
Then, for a given configuration of {\it projections} $\{\bR_a\}$, 
we may imagine {\it point particles\,} 
located at the set of positions $\{2\bR_a\}$,
as well as an extra point particle placed in the origin. 
These particles constitute a Poisson point process in the plane.
The Voronoi cell 
of the particle in the origin is then equal in distribution to the typical 
Poisson-Voronoi cell \cite{Moller94}, while also being identical to
the zero-cell of the $\alpha=2$ line tessellation.

Although not at present of any known application in physics,
it is instructive to study the entire $\alpha$ dependent family of problems
in order to see how it links together the two cases of greatest renown,
the Crofton cell for $\alpha=1$ and the typical
Poisson-Voronoi cell for $\alpha=2$. 

%%%%%%%%%%%%%%%%%%%%%%%%%%%%%%%%%%%%%%%%%%%%%%%%%%%%%%%%%%%%%%%%%%%%%%%%%%%%%%

\subsection{Results}
\label{secresults}

For the $\alpha$ dependent tessellation defined by (\ref{defrho})
we construct the large-$n$ expansion of the 
probability $p_n(\alpha)$ that the zero-cell is $n$-sided.
In section \ref{secfamily} we consider the parameter
range $\alpha>1$, for which our work is a rather straightforward extension
of earlier work \cite{Hilhorst05b} on the Voronoi cell, $\alpha=2$.
The final result for $p_n(\alpha)$ is best expressed with the aid
of the auxiliary quantity $p_n^{(0)}(\alpha)$ given by
\beq
{p}_n^{(0)}(\alpha) = \frac{2\,(4\pi^2\alpha)^{n-1}}{(2n)!}\,,
\qquad \alpha\geq 1. 
\label{resultpn0}
\eeq
In section \ref{secfamily} we derive that when $n\to \infty$, 
\beq
p_n(\alpha) \,{\simeq}\,
C(\alpha)\,p_n^{(0)}(\alpha), 
\qquad \alpha>1,
\label{resultpnalpha}
\eeq
where the symbol $\simeq$ is defined by the classical equivalence 
($u_n\simeq v_n$ if and only if ${u_n}/{v_n}{\to} 1$ for $n\to\infty$), and
where the prefactor $C(\alpha)$ is given by
\beq
C(\alpha) = \prod_{q=1}^\infty\, 
\Bigg( 1 - \frac{3-\alpha}{q^2} + \frac{\alpha^2}{q^4} \Bigg)^{-1},
\qquad \alpha>1.
\label{resultCalpha}
\eeq
Equivalently, we have that when $n\to \infty$, 
\beq
\log p_n(\alpha) {=}
-2n\log n+n\log(\pi^2{\rm e}^2\alpha)-\tfrac{1}{2}\log n +
\log\left(\frac{C(\alpha)}{4\pi^{5/2}\alpha}\right)+o(1).
\eeq
For $\alpha=2$ this result reduces to the known expressions 
\cite{Hilhorst05a,Hilhorst05b} of the Voronoi problem.
The study for $\alpha>1$ shows, however, that the Crofton cell 
corresponds to a singular point
at the lower limit, $\alpha=1$, of the parameter range studied here.
We have $\lim_{\alpha\to 1}C(\alpha)=\infty$, the divergence being due to
the factor of index $q=1$ in (\ref{resultCalpha}).
This signals the breakdown of the method of section \ref{secfamily}  
when applied to the Crofton cell, $\alpha=1$.

In section \ref{secCrofton} we 
develop the modified approach necessary
to find the sidedness probability $p_n(1)$ of the Crofton cell,
as well as its other statistical properties.
The appropriate result for the Crofton cell derived there is
that for $n\to\infty$
\beq
p_n(1) \simeq \tfrac{2}{3} n p_n^{(0)}(1), 
\label{resultpn1}
\eeq
or equivalently
\beq
\log p_n(1){=}
-2n\log n+n\log(\pi^2{\rm e}^2)+\tfrac{1}{2}\log n
-\log(6\pi^{5/2}) + o(1).
\label{resultpn1log}
\eeq 
As a corollary we find in section \ref{sectypical} that
the {\it typical\,} cell in a Poisson line tessellation
has a sidedness probability $p_n^{\rm typ}(1)$ given by
\beq
p_n^{\rm typ}(1) \simeq \tfrac{8}{3}n^{-1}p_n^{(0)}(1), \qquad n\to\infty,
\label{resultpntyp1}
\eeq
The ratio $p_n(1) / p_n^{\rm typ}(1) \simeq \tfrac{1}{4}n^2$  is
that of the area of the $n$-sided Crofton cell
to the area of the average cell in a Poisson line tessellation.

The derivation of these asymptotic series for the $p_n(\alpha)$ 
is based on a perturbation
expansion around the regular $n$-sided polygon centered at the origin.
The prefactor $C(\alpha)$ represents the partition function of the elastic
deformations of this $n$-gon, the `elasticity' being of purely
entropic origin. The deformations with $q=1$ are, at least to linear
order, translations of the cell with respect to the origin.
As is briefly discussed in section \ref{secgeneralproperties}, 
the expansion actually leads to the full probability density functional,
valid in the limit $n\to\infty$, of the zero-cell perimeter.

The possibility of constructing an asymptotic series by expanding around the
regular $n$-gon implies that when $n\to \infty$, 
the shape of the $n$-sided cell tends with probability one to a 
circle (which we show to be of radius $R_{\rm  c}=(n\alpha/2\pi)^{1/\alpha}$).
Miles \cite{Miles95} was the first to state this property 
for the Crofton cell and to validate it by heuristic methods.

The approach to circularity in the limit of large {\it sidedness\,} $n$ as
studied here resembles, 
but is nevertheless distinct from,
many mathematical results derived in the limit
of large cell {\it size,} with the size being defined in various ways.
These two limits correspond to ensembles that are fully disjoint.

The conjecture originally
formulated by Kendall in the early forties says that the Crofton
cell becomes circular when its {\it area\,} goes to infinity. 
Calka and Schreiber \cite{CalkaSchreiber05}
proved the approach to a circle when the radius of
the largest possible {\it inscribed disk\,} tends to infinity.
Hug, Reitzner and Schneider  \cite{Hugetal04} 
showed a generalized version of Kendall's conjecture in any
dimension $d$ when the $k$-dimensional volume  of the cell, 
in the sense of Hausdorff ($2\le k\le d$), goes to infinity.
Hug et Schneider \cite{HugSchneider07} considered the zero-cell of a general
class of $d$-dimensional tessellations which includes the
two-dimensional family studied in this work, 
but also includes cells (`polytopes') resulting from more general 
direction dependent Poisson hyperplane processes.
These authors prove the approach to a limit shape when
the size of the cell, as measured in any of a variety of
ways, tends to infinity.

The present work, in summary, considers a 
special subclass of the systems introduced in reference \cite{HugSchneider07}
and distinguishes itself from that work in that it
takes a different large-cell limit and leads to an explicit
expansion for the sidedness probability. 
Our expansion in negative powers of $n$
is not mathematically rigorous;
but it is of a kind that in statistical mechanics
commonly leads to exact results.

%%%%%%%%%%%%%%%%%%%%%%%%%%%%%%%%%%%%%%%%%%%%%%%%%%%%%%%%%%%%%%%%%%%%%%%%%%%%
%%%%%%%%%%%%%%%%%%%%%%%%%%%%%%%%%%%%%%%%%%%%%%%%%%%%%%%%%%%%%%%%%%%%%%%%%%%%
%%%%%%%%%%%%%%%%%%%%%%%%%%%%%%%%%%%%%%%%%%%%%%%%%%%%%%%%%%%%%%%%%%%%%%%%%%%%

\section{A one-parameter family of line tessellations}
\label{secfamily}

%%%%%%%%%%%%%%%%%%%%%%%%%%%%%%%%%%%%%%%%%%%%%%%%%%%%%%%%%%%%%%%%%%%%%%%%%%%%

We consider a set of lines defined by projections $\bR$
that are distributed in the plane ${\mathbb R}^2$ with density
\beq
\rho(\bR)=\lambda R^{\alpha-2}, \qquad  R>0, \quad \alpha \geq 1, 
\label{defrhobis}
\eeq
where $\lambda$ has the dimension of an inverse 
length to the power
$\alpha$ and will be kept only as a check on the dimensionalities 
of our formulas.
The expected number ${\cal N}(\alpha,L)$ of projections
in a disk of radius $L$ centered at the origin is equal to
\beq
{\cal N}(\alpha,L) = \int_{R<L}\dd\bR\,\rho(\bR) 
                   = \frac{2\pi\lambda L^{\alpha}}{\alpha}.
\label{rescalN}
\eeq
In order to have a well-defined problem in the infinite plane 
we first consider this disk occupied by $N$ 
projections distributed
independently according to the probability density
\beq
P(\bR)= \frac{\alpha}{2\pi L^2} \Big(\frac{L}{R}\Big)^{2-\alpha},  
        \qquad   0<R<L.
\label{defP}
\eeq
At some suitable point below we will take the limit $L,N\to\infty$ with 
$N = {\cal N}(\alpha,L)$.

%%%%%%%%%%%%%%%%%%%%%%%%%%%%%%%%%%%%%%%%%%%%%%%%%%%%%%%%%%%%%%%%%%%%%%%%%%%%%

\subsection{Sidedness probability $p_n(\alpha)$ of the zero-cell}
\label{seccellorigin}

Each projection $\bR_a$, for $a=1,2,\ldots,N$, has a line $a$ 
associated with it.
Then $p_n$ is the probability that $n$ lines from among the $N$ contribute a
segment to the perimeter of the zero-cell (that is, the cell
enclosing the origin), {\it and} that
the other $N-n$ lines do not contribute segments. We will
labeling these two groups of lines by $a=1,2,\ldots,n$ and 
$a=n+1,n+2,\ldots,N$, respectively, 
and abbreviate $\bRn=\{\bR_1,\ldots,\bR_n\}$.
Omitting explicit indication of
the dependence of $p_n(\alpha)$ on $N$ and $L$,
we can write 
\bea
p_n(\alpha) &=& \binom{N}{n} \int_{R_1<L}\dd\bR_1P(\bR_1)\ldots
                     \int_{R_n<L}\dd\bR_nP(\bR_n)\, \chi(\bRn)
\nonumber\\[2mm]
&& \times \exp\big( -{\cal A}_\alpha(\bRn) \big),
\label{defpn}
\eea
where $\chi$ selects the $\bRn$ that define Crofton cells: 
\beq
\chi(\bRn)=
\left\{
\begin{array}{ll}
1 & \mbox{if $\bRn$ represents an $n$-sided cell enclosing the origin,}
\\[2mm] 
0 & \mbox{in all other cases};
\end{array}
\right.
\label{defchi0G}
\eeq
and $\exp\!\big( -{\cal A}_\alpha(\bRn) \big)$ 
is the probability that the other $N-n$ lines
do not intersect the perimeter of this cell. 
We will employ polar coordinates and write $\bR_a=(R_a,\Phi_a)$.
Upon using (\ref{defP}) for $P$
and partially taking the limit $N\to\infty$ we can rewrite (\ref{defpn}) as
\beq
{p}_n(\alpha) = \frac{\lambda^n}{n!}
\int_0^{2\pi}\dd\Phi_1\ldots\dd\Phi_n 
\int_0^\infty \dd R_1 R_1^{\alpha-1}  \ldots  \dd R_n R_n^{\alpha-1}  
\, \chi(\bRn)\,\ee^{-{\cal A}_\alpha(\bRn)}.
\label{defpnbis}
\eeq
In order to make progress
we must now render ${\cal A}_\alpha$ and $\chi$ explicit.

%%%%%%%%%%%%%%%%%%%%%%%%%%%%%%%%%%%%%%%%%%%%%%%%%%%%%%%%%%%%%%%%%%%%%%%%%%%%%%

\subsubsection{Expression for ${\cal A}_\alpha$}
\label{secsupportfunction}

\begin{figure}
\begin{center}
\scalebox{.55}
{\includegraphics{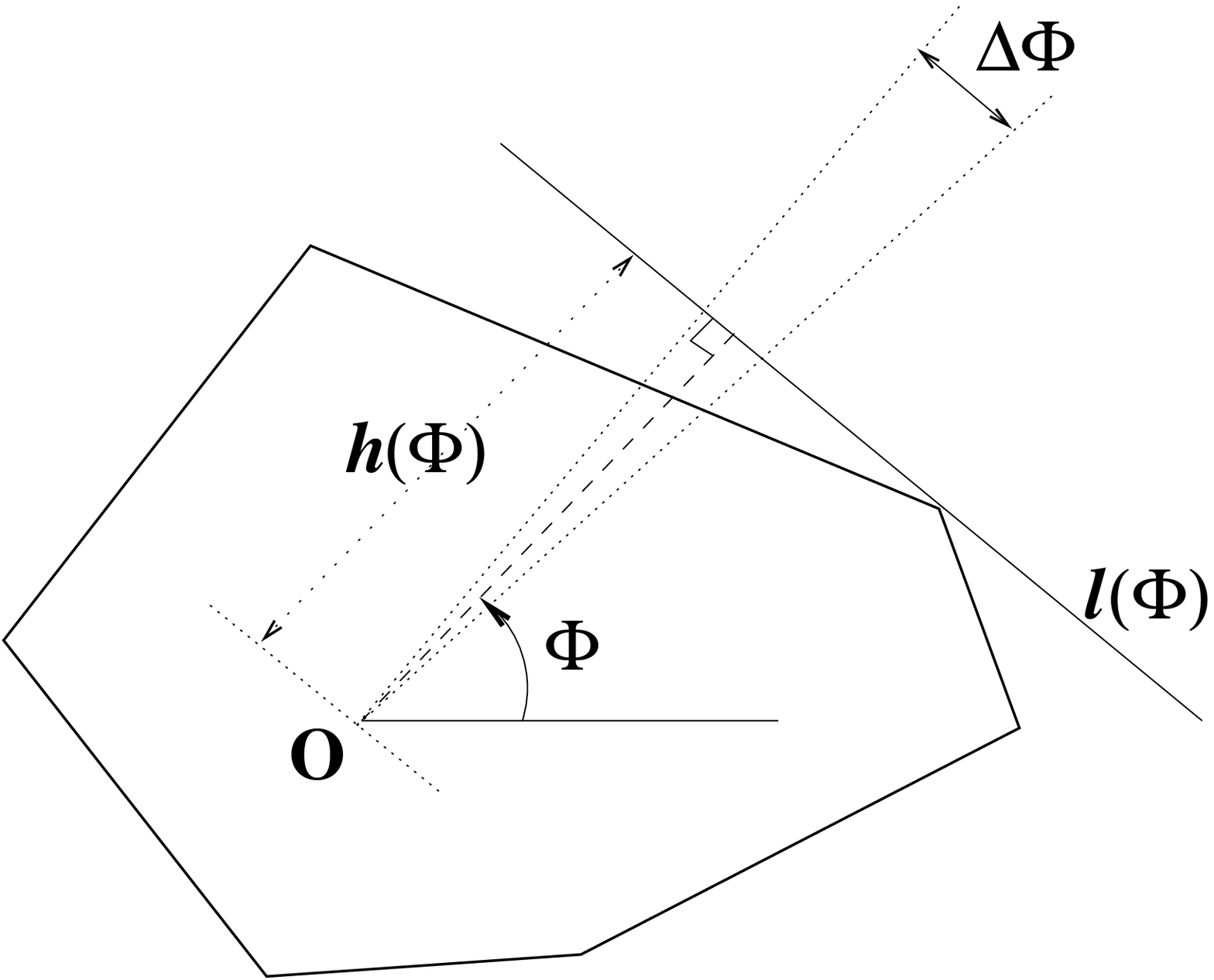}}
\end{center}
\caption{\small
Figure illustrating the definition of the support function $h(\Phi)$
given in section \ref{secsupportfunction}.}
\label{hPhi}
\end{figure}

Since the $N-n$ lines are independent, $\exp(-{\cal A}_\alpha)$ is an
$(N-n)$th power and we need to consider only a single line.
We refer now to figure \ref{hPhi}.
%%%%%%%%%%%%%%%%%%%%%%%%%%%%%%%%
%%%%%%%%%%%%%%%%%%%%%%%%%%%%%%%%
%%%%%%%%%%%%%%%%%%%%%%%%%%%%%%%%
%%%%%%%%%%%%%%%%%%%%%%%%%%%%%%%%
For a half-line originating in the origin ${\bf O}$ 
and having an arbitrary angular direction $\Phi$, let
$\ell(\Phi)$ be the tangent to the cell that intersects this half-line 
perpendicularly. 
Generically this tangent will have only
a single vertex in common with the perimeter.
The distance $h(\Phi)$ from the origin to $\ell(\Phi)$ 
is called the `support function' \cite{Schneider93} of $\Phi$.
The no-intersection condition can now be restated as the condition that there
be no projection $\bR_a$ ($a=n+1,\ldots,N$)
in the sector of width $\Delta\Phi$
within a distance $h(\Phi)$ of the origin. The probability 
$w(\Phi)\Delta\Phi$ for a given $\bR_a$ to be in this sector is
\bea
w(\Phi)\Delta\Phi &=& \int_{\Phi}^{\Phi+\Delta\Phi} \dd\Phi'\!
                      \int_0^{h(\Phi')}\! \dd R\,R\,P(\bR) \nonumber\\[2mm]
&=& \frac{\Delta\Phi}{2\pi}\Big(\frac{h(\Phi)}{L}\Big)^{\alpha} 
    + {\cal O}(\Delta\Phi^2),
\label{resw}
\eea
where we used (\ref{defP}).
The probability $\exp(-{\cal A}_\alpha)$
for the line through $\bR_a$ not to intersect the cell
is equal to one minus the integral on all $\Phi$ of expression (\ref{resw}).
Upon raising this result to the $(N-n)$th power and taking the limit
$N\to\infty$ we obtain the desired weight
$\exp(-{\cal A}_\alpha)$ with ${\cal A}_\alpha$ given by
\beq
{\cal A}_\alpha(\bRn) = \frac{\lambda}{\alpha}
\int_0^{2\pi}\dd\Phi\, h^{\alpha}(\Phi).
\label{defcalAalpha}
\eeq
Equations (\ref{defpnbis}) and (\ref{defcalAalpha}) 
can also be obtained from Slivnyak's formula by
the method of reference \cite{Calka03a}. Formula (\ref{defpnbis}) was
originally included in \cite{MilesMaillardet82}. 
Whenever (\ref{defcalAalpha}) appears below,
we will set $\lambda=1$, which entails no loss of generality.

It is easy to find
the explicit expression for the support function $h(\Phi)$, 
which follows directly from its above definition. 
By a permutation of the labels of the polar angles and a rotation
of the coordinate system one can always arrange things such that 
$0=\Phi_0<\Phi_1<\ldots<\Phi_{n-1}<2\pi$,
where by convention $\Phi_n\equiv\Phi_0$.
We define the ${\bS_m}=(S_m,\Psi_m)$ as the
consecutive vertices of the cell in polar coordinate representation;
in particular, ${\bS_m}$ is the intersection of the
$(m-1)$-th with the $m$-th line (see figure \ref{figangles}).
After that a little algebra yields the two alternative expressions 
\bea
h(\Phi) &=& \big[ \sin(\Phi_m-\Phi_{m-1}) \big]^{-1}
\big[ R_{m-1}\sin(\Phi_m-\Phi) + R_{m}\sin(\Phi-\Phi_{m-1}) \big] 
\nonumber\\[2mm]
       &=& S_m \cos(\Psi_m-\Phi),
\qquad \Phi_{m-1}<\Phi<\Phi_m, \quad m=1,\ldots,n
\label{defh}
\eea 
When one substitutes (\ref{defh}) in (\ref{defcalAalpha}),
a $\Phi$ integral appears which can be carried out in closed form 
only for integer $\alpha$. However,
the large-$n$ expansion to be introduced in the course of our 
development will apply for arbitrary $\alpha$.

%%%%%%%%%%%%%%%%%%%%%%%%%%%%%%%%%%%%%%%%%%%%%%%%%%%%%%%%%%%%%%%%%%%%%%%%%%%%

\subsubsection{Angular variables}
\label{secangles}

\begin{figure}
\begin{center}
\scalebox{.60}
{\includegraphics{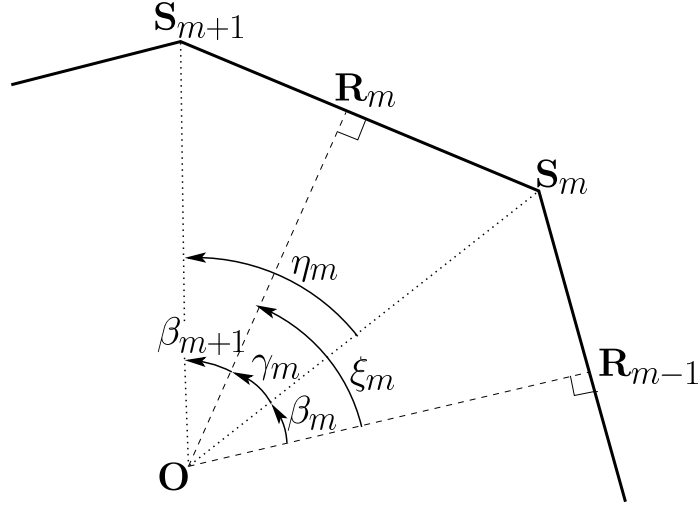}}
\end{center}
\caption{{\small
Heavy line segments: the perimeter of the zero-cell
in a random line tessellation of the plane.
The $\bR_m$ are the projections of the origin onto the lines containing
these perimeter segments.
The $\bS_m$ are the cell vertices.
The figure shows the angles $\xi_m$, $\eta_m$, $\beta_m$, and $\gamma_m$
defined by (\ref{defxieta}) and (\ref{inversebgxy}).
}} 
\label{figangles}
\end{figure}

Other angles essential to our study may be defined in terms of the 
$\Phi_m$ and $\Psi_m$ and are shown in figure \ref{figangles}.
First of all,
\bea
\xi_m &=& \Phi_m-\Phi_{m-1}\,, \nonumber\\[2mm]
\eta_m &=& \Psi_{m+1}-\Psi_m, \qquad m=1,2,\ldots,n,
\label{defxieta}
\eea
with the conventions $\Phi_{n}=\Phi_0+2\pi$
and $\Psi_{n}=\Psi_0+2\pi$.
The $\xi_m$ are the angles between two consecutive projection
vectors and the $\eta_m$ those between two consecutive vertex vectors; 
$n$-periodicity in their index $m$ will be understood.
For fixed sets of angles
$\xi=\{\xi_m\}$ and $\eta=\{\eta_m\}$ one may still jointly rotate
the vertex vectors $\bS_m$ with respect to the projection vectors
$\bR_m$,
as this modifies only the relative angles $\beta_m$ and $\gamma_m$ 
(see figure \ref{figangles}) between the two sets. 
We may select any one of these relative angles and call it `the' angle of
rotation, since it will determine all others. We will select $\beta_1$
for this purpose;
when it is given, the remaining $\beta_m$ and $\gamma_m$ can be
expressed as
\bea
\beta_m &=& \phantom{-}\beta_1 - 
\sum_{\ell=1}^{m-1}(\xi_{\ell}-\eta_\ell),
\qquad m=2,\ldots,n,
\nonumber\\
\gamma_m &=& -\beta_1 +
\sum_{\ell=1}^{m-1}(\xi_{\ell}-\eta_{\ell}) + \xi_m\,, 
\qquad m=1,\ldots,n,
\label{inversebgxy}
\eea
We now show that $\beta_1$ cannot be arbitrary but is in fact determined
itself by the two sets $\xi$ and $\eta$.
As is clear from figure \ref{figangles}, one can relate
$R_{m}$ to $R_{m-1}$ by 
$R_{m}=(\cos\gamma_m/\cos\beta_m)R_{m-1}$.
Upon iterating $n$ times starting from any of the $R_m$, 
periodicity imposes that we recover the initial value.
Let us define $G$ by
\beq
\ee^{2\pi G(\xi,\eta;\beta_1)} = 
\prod_{m=1}^n \frac{\cos\gamma_m}{\cos\beta_m}\,,
\label{defG}
\eeq
where the notation expresses that the $2n-1$ variables
$\gamma_1,\beta_2,\gamma_3,\ldots,\beta_n$ on the right hand side
should be viewed as the functions (\ref{inversebgxy}) 
of the $\xi_m$ and $\eta_m$ and of the `angle of rotation' $\beta_1$.
Because of periodicity $\beta_1$ must then have the special value
$\beta_1=\beta_*(\xi,\eta)$ that is the solution of
\beq
G(\xi,\eta;\beta_*)=0.
\label{nospiralcond}
\eeq
It was shown in reference \cite{Hilhorst07} that the solution of
(\ref{nospiralcond}) is unique.

Finally we remark that whereas the $\xi_m$ and $\eta_m$ must
be positive, the $\beta_m$ and $\gamma_m$ may have either sign.
\vspace{3mm}

In terms of the angles defined above,
equations (\ref{defcalAalpha})-(\ref{defh}) can be made
fully explicit in those cases where it is possible
to carry out the integral on $\Phi$. One such case occurs for
$\alpha=1$. By Cauchy's integral formula we have
${\cal A}_1={\cal P}$, 
where ${\cal P}$ is the cell perimeter and 
has the two alternative expressions 
\bea
{\cal P} &=& \summ R_m \big[ \tan\tfrac{1}{2}\xi_m 
                       +\tan\tfrac{1}{2}\xi_{m+1} \big]
\nonumber \\
         &=& \summ R_m \big[ \tan\gamma_m + \tan\beta_{m+1} \big].
\label{defcalP}
\eea
For $\alpha=2$ 
we have that ${\cal A}_2$ is the area of the fundamental domain of the 
Voronoi cell, that is, of the union of $n$ disks of radii $S_m$ 
centered at the vertices $\bS_m$.
Explicit expressions for ${\cal A}_2$ may be found in references
\cite{Calka03a,Calka03b,Hilhorst05a,Hilhorst05b,Hilhorst07}.

%%%%%%%%%%%%%%%%%%%%%%%%%%%%%%%%%%%%%%%%%%%%%%%%%%%%%%%%%%%%%%%%%%%%%%%%%%%%

\subsubsection{Expression for the indicator function $\chi$}
\label{secchi}

Having defined the angular variables
we return now to the conditions imposed on the domain of integration
in (\ref{defpnbis})
by the indicator $\chi$. 
In terms of the angles $\beta_m$ and $\gamma_m$ these conditions
simplify greatly and take the explicit form \cite{Hilhorst05a,Hilhorst05b} 
\beq
-\tfrac{\pi}{2} < \beta_m, \gamma_m < \tfrac{\pi}{2},  \qquad
\beta_m+\gamma_m > 0, \qquad \gamma_m+\beta_{m+1}>0,
\label{defchi}
\eeq
where $m=1,2,\ldots,n$ and $n$-periodicity in $m$ is understood.
As could be expected,
this condition depends only on the angles of the problem 
and not on the length scale.

%%%%%%%%%%%%%%%%%%%%%%%%%%%%%%%%
%%%%%%%%%%%%%%%%%%%%%%%%%%%%%%%%
%%%%%%%%%%%%%%%%%%%%%%%%%%%%%%%%
%%%%%%%%%%%%%%%%%%%%%%%%%%%%%%%%

%%%%%%%%%%%%%%%%%%%%%%%%%%%%%%%%%%%%%%%%%%%%%%%%%%%%%%%%%%%%%%%%%%%%%%%%%%%%%
%%%%%%%%%%%%%%%%%%%%%%%%%%%%%%%%%%%%%%%%%%%%%%%%%%%%%%%%%%%%%%%%%%%%%%%%%%%%%
%%%%%%%%%%%%%%%%%%%%%%%%%%%%%%%%%%%%%%%%%%%%%%%%%%%%%%%%%%%%%%%%%%%%%%%%%%%%%

\subsection{Transforming the expression for $p_n(\alpha)$}
\label{sectransforming}

Equation (\ref{defpnbis}) represents $p_n(\alpha)$ as a $2n$-fold integral
which we will transform in successive steps to a form manageable
in the large $n$ limit. An important variable is the `average radius'
$\Rav$ defined by
\beq
R_{\rm av} = \frac{1}{n}\summ R_m.
\label{defRavrhom}
\eeq
We now pass from the set of variables of integration
$\{R_m,\Phi_m\}$ used in (\ref{defpnbis}) to a new set of variables 
consisting of the radial scale $R_{\rm av}$ and the set of angles 
$\{\xi_m,\eta_m\}$. 
Although the $\beta_m$ and $\gamma_m$ may be entirely expressed in terms
of the $\xi_\ell$ and $\eta_\ell$, they constitute a useful set
of auxiliary variables.  

Convenient shorthand notation will be
\beq
\int_{\xi,\eta} \equiv \int_0^{2\pi}\!\dd\xi_1\,\xi_1 \ldots \dd\xi_n\,\xi_n
                    \int_0^{2\pi}\!\dd\eta_1\ldots\dd\eta_n\,
\delta\big(\sum_{m=1}^n\xi_m-2\pi\big)\,
\delta\big(\sum_{m=1}^n\eta_m-2\pi\big) 
\label{defintxieta}
\eeq
and
\beq
\Theta \equiv \prod_{m=1}^n \theta(\tfrac{\pi}{2}-\beta_m)
         \prod_{m=1}^n \theta(\tfrac{\pi}{2}-\gamma_m),
\label{defTheta}
\eeq
where $\theta$ is the Heaviside unit step function.
After doing the algebra of the transformation of
variables \cite{Hilhorst05b,Hilhorst07} 
we find that (\ref{defpnbis}) may be recast in the form
\beq
p_n(\alpha) = \frac{1}{n} \int_{\xi,\eta}
G'(\xi,\eta;\beta_*)^{-1}
\Big[\prod_{m=1}^n \rho_m^\alpha T_m^{\phantom{2}}\xi_m^{-1} \Big] 
\,\Theta\,\int_0^\infty \!\dd R_{\rm av}\, R_{\rm av}^{\alpha n -1}\,\,
\ee^{-{\cal A}_\alpha},
\label{pnbeforeint}
\eeq
in which $G'$ denotes the derivative of $G(\xi,\eta;\beta_1)$ with respect to
$\beta_1$; we have abbreviated
$T_m \equiv \sin\xi_m/\cos^2\beta_m$; and 
the $2n$ ratios $\rho_m \equiv R_m/R_{\rm av}$ may be expressed in terms
of the angles by means of the $2n$ relations
\beq
\rho_m = (\cos\gamma_m / \cos\beta_m) \rho_{m-1}\quad (m=2,\ldots,n),
\qquad n^{-1}\summ\rho_m = 1,
\label{rhom}
\eeq
of which the last one follows from (\ref{defRavrhom}).

Clearly ${\cal A}_\alpha$ can be written as $R_{\rm av}^{\alpha}$
times a function of the angles. We will set
\beq
{\cal A}_\alpha = \frac{2\pi}{\alpha}\,R_{\rm av}^{\alpha}(1 + n^{-1}V),
\label{exprcalAalpha}
\eeq
where $2\pi\alpha^{-1}R_{\rm av}^{\alpha}$ is the value that 
${\cal A}_\alpha$
takes for a circular cell of radius $R_{\rm av}$ and in which $V$ depends
exclusively on the angles. The notation in (\ref{exprcalAalpha}) is meant to 
suggest that $n^{-1}V$ is of order $n^{-1}$ as $n\to\infty$,
but we will rely on that only later. 
Integrating (\ref{pnbeforeint}) on $R_{\rm av}$ converts it into 
\beq
p_n(\alpha) =
\frac{(n-1)!}{2\pi n} \Big( \frac{\alpha}{2\pi} \Big)^{n-1} 
\int_{\xi,\eta} \Theta\,\ee^{-{\mathbb V}(\alpha)}\,,
\qquad \alpha \geq 1,
\label{pnafterint}
\eeq
where
\beq
\ee^{-{\mathbb V}(\alpha)} = G'(\xi,\eta;\beta_*)^{-1}
\Big[ \prod_{m=1}^n \rho_m^\alpha T_m^{\phantom{2}} \xi_m^{-1} \Big]
(1+n^{-1}V)^{-n}.
\label{defmathbbV}
\eeq
The integrations that remain on the right hand side of (\ref{pnafterint})  
bear only on the angles $\xi_m$ and $\eta_m$, 
that is, on the shape of the cell without regard to its radial dimension.
We still remark that
the factors $\xi_m^{-1}$ in the product on $m$ in (\ref{defmathbbV})
compensate the factors $\xi_m$ incorporated 
in the definition (\ref{defintxieta}) of $\int_{\xi,\eta}$. 
Their purpose is to ensure that ${\mathbb V}(\alpha)$
remains finite when any of the $\xi_m$ tends to zero.

We finally write (\ref{pnafterint}) as
\beq
p_n(\alpha) = p_n^{(0)}(\alpha) \la\Theta\ee^{-{\mathbb V(\alpha)}} \ra,
\qquad \alpha \geq 1,
\label{pnfactors}
\eeq
where for any function $X$ of the angular variables we define
\beq
\la X \ra = \frac{ \int_{\xi,\eta} X }
                 { \int_{\xi,\eta} 1 }\,.
\label{deflara}
\eeq
Straightforward calculation yields
\bea 
p_n^{(0)}(\alpha) &=& 
\frac{(n-1)!}{2\pi n} \Big( \frac{\alpha}{2\pi} \Big)^{n-1}  
\int_{\xi,\eta} 1
\nonumber\\[2mm]
&=&
2\,\frac{ (4\pi^2\alpha)^{n-1} }{(2n)!}\,,  
\qquad \alpha \geq 1,
\label{calcpn0}
\eea
which is equation (\ref{resultpn0}) of the introduction. 
Equations (\ref{pnfactors}) and (\ref{calcpn0}) represent an important step
forward with respect to the initial expressions 
(\ref{defpnbis}) and (\ref{defcalAalpha}).
However, the hard part of the problem remains, namely to determine
the $n$ dependence of $\la\Theta\ee^{-{\mathbb V(\alpha)}}\ra$ 
in equation $(\ref{pnfactors})$.
In sections \ref{secexpansion}-\ref{secexppnalpha} 
we will show that for $\alpha>1$ 
$\lim_{n\to\infty} \la\Theta\ee^{-{\mathbb V(\alpha)}}\ra = C(\alpha)$, 
where $C(\alpha)$ is a finite numerical constant.
The special case of the Crofton cell, $\alpha=1$, will not be covered by
the arguments of those sections 
and must be dealt with separately; we will do so 
in section \ref{secCrofton}.

%%%%%%%%%%%%%%%%%%%%%%%%%%%%%%%%%%%%%%%%%%%%%%%%%%%%%%%%%%%%%%%%%%%%%%%%%%%%

\subsection{Scaling of the average radius $\Rav$ }
\label{secRav}
%%% [*CR 1011]

The inner integrand of (\ref{pnbeforeint}),
when combined with (\ref{exprcalAalpha}), shows that the average radius
$\Rav$ has a probability distribution 
\beq
P_{\rm av}(\Rav) = \mbox{cst}\times\Rav^{\alpha n-1}\,
\ee^{-(2\pi/\alpha)\Rav^\alpha(1+n^{-1}V)},
\qquad \alpha \geq 1,
\label{defPavRav}
\eeq
which depends on the cell shape through $V$.
The notation is meant to suggest that 
$n^{-1}V$ is negligible when $n\to\infty$, and this will be confirmed 
below. 
Elementary analysis then shows that for $n\to\infty$ the distribution
$P_{\rm av}(\Rav)$ has a peak at $\Rc$ whose width is $\sigma_{\rm c}$,
where
\beq
\Rc = \Big( \frac{n\alpha}{2\pi} \Big)^{\frac{1}{\alpha}}, \qquad
\sigma_{\rm c} = 
\frac{1}{\alpha}\Big( \frac{\alpha}{2\pi} \Big)^{\frac{1}{\alpha}}
n^{\frac{1}{\alpha}-\frac{1}{2}}.
\label{defRcsigmac}
\eeq
Hence the typical deviation of $\Rav$ from  
$R_{\rm c}$ is a factor $n^\half$ smaller
than $R_{\rm c}$ itself, for all $\alpha \geq 1$.

%%%%%%%%%%%%%%%%%%%%%%%%%%%%%%%%%%%%%%%%%%%%%%%%%%%%%%%%%%%%%%%%%%%%%%%%%%%%%

\subsection{Large-$n$ expansion of ${\mathbb V}(\alpha)$}
\label{secexpansion}

The regular $n$-sided polygon is a point of high symmetry in phase space
where we expect the integrand of $\int_{\xi,\eta}$ in (\ref{pnafterint})
to be stationary.
This point has $\xi_m=\eta_m=2\pi n^{-1}$
and $\beta_m=\gamma_m=\pi n^{-1}$ for all $m=1,2,\ldots,n$.
Coordinates that describe
the deviations from this symmetric state are defined by
\beq
\delta\xi_m = \xi_m - 2\pi n^{-1}, \qquad
\delta\eta_m = \eta_m - 2\pi n^{-1}
\label{defdeltaxieta}
\eeq
and their suitably scaled Fourier transforms
\beq
\hat{X}_q= \frac{n^\half}{2\pi}\summ \ee^{2\pi{\rm i} qm/n} \delta\xi_m\,,
\qquad 
\hat{Y}_q= \frac{n^\half}{2\pi}\summ \ee^{2\pi{\rm i} qm/n} \delta\eta_m\,, 
\label{defXqYq}
\eeq
where $q=\pm 1,\pm 2,\ldots$. It will appear that 
the integrations required for the calculation of the average in
(\ref{pnfactors}) can be done explicitly once expressed in terms of the
variables of integration $\hat{X}_q$ and $\hat{Y}_q$.

%%%%%%%%%%%%%%%%%%%%%%%%%%%%%%%%%%%%%%%%%%%%%%%%%%%%%%%%%%%%%%%%%%%%%%%%%%%%%%%

\subsubsection{Scaling with $n$}
\label{secscaling}

We will treat ${\mathbb V}$ perturbatively in inverse powers of $n$. 
An initial hypothesis on the smallness
of the angles with $n$ is suggested by the scaling that prevails when in
(\ref{pnafterint}) we set ${\mathbb V}=0$. That yields
\beq
\delta\xi_m,\;\delta\eta_m \sim {n}^{-1},
\label{hypo}
\eeq
where the symbol $\sim$ indicates the scaling with $n$ in the large-$n$
limit. 
Using relations (\ref{inversebgxy}) and (\ref{rhom}) between the angles 
as well as the fact that 
the $\xi_m$ and $\eta_m$ are to leading order independent,
we find from (\ref{hypo}) that furthermore
\beq
\beta_m,\gamma_m\sim n^{-\half},
\qquad \tau_m \equiv \rho_m-1 \sim n^{-\half};
\label{hypo2}
\eeq
and using (\ref{defXqYq}) we find from (\ref{defdeltaxieta})
that
\beq
\hat{X}_q \sim n^0, \qquad \hat{Y}_q \sim n^0.
\label{hypo3}
\eeq
If and when needed, the $\beta_m$, $\gamma_m$, and $\tau_m$ and their Fourier
transforms can be expressed in terms of the $\hat{X}_q$ and $\hat{Y}_q$.

The initial hypothesis now holds that the above scalings remain valid in 
the presence of the nonzero ${\mathbb V}$ defined by (\ref{defmathbbV}).
We will consider the validity of this hypothesis confirmed 
when at the end of this section we will find that 
in the limit $n\to\infty$ it produces a finite leading order result for
$\la\exp(-{\mathbb V})\ra$.

We end these considerations with two remarks.
First, the present scalings are identical to those encountered in the study
of the large $n$-sided Voronoi cell. How to handle them technically, in
particular when sums of $\sim\!n$ terms appear, is nontrivial and
has been discussed
in detail in reference \cite{Hilhorst05b}, sections 5 and 6.
Secondly, we note that the smallness of the $\tau_m$ with 
$n$ implies that all $R_m$ are close to the average $\Rav$, which in turn
together with the smallness of the $\xi_m$ implies that for $n\to\infty$ the 
perimeter approaches a circle.

%%%%%%%%%%%%%%%%%%%%%%%%%%%%%%%%%%%%%%%%%%%%%%%%%%%%%%%%%%%%%%%%%%%%%%%%%%%%%

\subsubsection{The expansion}
\label{secexpansionsub}
 
We defined ${\mathbb V}$ recursively
by (\ref{defmathbbV}), (\ref{exprcalAalpha}), and (\ref{defcalAalpha}) 
in terms of $V$, ${\cal A}_\alpha$, and $h(\Phi)$.
Since $h(\Phi)$ defined by (\ref{defh}) is piecewise analytic on $n$
successive angular intervals,
when substituted in (\ref{defcalAalpha}) it gives rise to a sum of $n$ terms.
In the $m$th term, that is, for $\Phi_{m-1} < \Phi < \Phi_m$,
we set $S_m = R_m/\!\cos\gamma_m$ and pass to the variable of integration
$\theta\equiv\Phi-\Psi_m$. 
Then (\ref{defcalAalpha}) becomes
\bea 
{\cal A}_\alpha &=& \frac{1}{\alpha} \summ 
  \Big( \frac{R_m}{\cos\gamma_m} \Big)^\alpha
  \int_{-\beta_m}^{\gamma_m}\!\dd\theta\,(\cos\theta)^\alpha
\nonumber\\[2mm]
&=& \frac{\Rav^\alpha}{\alpha} \summ
 \Big( \frac{1+\tau_m}{1-\half\gamma_m^2+\ldots} \Big)^\alpha
   \int_{-\beta_m}^{\gamma_m}\!\dd\theta\,
   \big( 1-\tfrac{1}{2}\theta^2+\ldots \big)^\alpha.
\label{expansioncalA}
\eea
Here the dots indicate terms of higher order in the angles.
We may expand the $\alpha$th powers
in (\ref{expansioncalA}) and do the $\theta$ integral.
%%% [*CR 1003]
Using the sum rule $\summ(\beta_m+\gamma_m)=2\pi$ we find that the leading
order result for ${\cal A}_\alpha$ is $(2\pi/\alpha)\Rav^\alpha$.
Comparison to (\ref{exprcalAalpha}) 
shows that the higher order terms in (\ref{expansioncalA}) determine $V$.
Pursuing the expansion to higher orders we find
\bea 
\frac{2\pi}{n}V &=& \alpha\summ\tau_m(\gamma_m+\beta_m)
  +\tfrac{1}{2}\alpha(\alpha-1)\summ\tau_m^2(\gamma_m+\beta_m)
\nonumber\\[2mm]
&& +\,\tfrac{1}{6}\alpha ( 2\gamma_m^3 + 3\gamma_m^2\beta_m - \beta_m^3 )
+\ldots\,.
\label{expansionV}
\eea
Equation (\ref{expansionV}) for $V$ may be substituted in (\ref{defmathbbV}).
The other factors in (\ref{defmathbbV}) may be expanded similarly,
and together this leads to the expansion of ${\mathbb V}$.
The order in $n$ of each term in the expansion 
may be estimated in the way outlined in section \ref{secscaling}.
It then appears that ${\mathbb V}(\alpha)$  allows for an expansion 
in powers of $n^{-\half}$, 
\beq
{\mathbb V}(\alpha) = {\mathbb V}_1(\alpha) + {\cal O}(n^{-\half}),
\label{expmathbbV}
\eeq
of which the leading term ${\mathbb V}_1(\alpha)$ 
is a quadratic form in the angles that is of order $n^0$. 
Upon expressing all variables in terms of the $\hat{X}_q$ and $\hat{Y}_q$ 
one obtains
\beq
{\mathbb V}_1(\alpha) = \sum_{q\neq 0}\, 
(\hat{X}_q,\hat{Y}_q)\cdottt{\bsV}_q\cdottt
                     (\hat{X}_{-q},\hat{Y}_{-q})^{\rm T}\,,
\label{resV1}
\eeq
where the superscript {\sc T} indicates transposition and
where $\bsV_q$ is the symmetric matrix 
\beq
{\bsV}_q = \left(
\begin{array}{ll}
 \phantom{-}A_q      \phantom{xx} & -A_q+\tfrac{1}{2}B_q \\[2mm]
-A_q+\tfrac{1}{2}B_q \phantom{xx} & \phantom{-}A_q-B_q
\end{array}
\right)
\label{exprBq}
\eeq
with
\beq
A_q = \frac{\alpha-1}{q^2} + \frac{\alpha^2}{2q^4}\,,   \qquad    
B_q = \frac{\alpha}{q^2}\,.
\label{defAqBq}
\eeq
Equations (\ref{expmathbbV})-(\ref{defAqBq}) 
complete the large-$n$ expansion of ${\mathbb V}(\alpha)$.
They are valid for all $\alpha \geq 1$. 

%%%%%%%%%%%%%%%%%%%%%%%%%%%%%%%%%%%%%%%%%%%%%%%%%%%%%%%%%%%%%%%%%%%%%%%%%%%%

\subsection{Large-$n$ expansion of $p_n(\alpha)$}
\label{secexppnalpha}

The large-$n$ expansion of $p_n(\alpha)$ is based on the one of
${\mathbb V}$ given above.
Upon substituting (\ref{expmathbbV})-(\ref{resV1}) 
in (\ref{pnfactors}) and using that
$\Theta$ may be replaced with unity up to corrections that vanish
exponentially for large $n$, we get 
\beq
p_n(\alpha) \simeq p_n^{(0)}(\alpha) \la \ee^{-{\mathbb V}_1(\alpha)} \ra
\label{pnaspt}
\eeq
with ${\mathbb V}_1$ given by (\ref{resV1}).

Although (\ref{resV1}) is a quadratic form,
the $\hat{X}_q$ and $\hat{Y}_q$ are not Gaussian distributed. 
Nevertheless, it was shown in reference \cite{Hilhorst05b} that 
to leading order in $n^{-1/2}$ the average $\la\ldots\ra$ in
(\ref{pnaspt}) may be carried out as though the $\hat{X}_q$ and $\hat{Y}_q$ 
were Gaussian, with a probability distribution 
\beq
{\cal N}_G\,\exp\Big(\! -\sumqno
(\hat{X}_q,\hat{Y}_q)\cdottt{\bsE^{-1}}\cdottt
                     (\hat{X}_{-q},\hat{Y}_{-q})^{\rm T} \Big),
\label{Gaussianweight}
\eeq
where ${\cal N}_G$ is the appropriate normalization constant and
where we introduced the $2 \times 2$ diagonal matrix 
$\,\bsE = {\rm diag}\{1,2\}$. This fact is partly confirmed by the result of
convergence in distribution and asymptotic mutual independence of the Fourier
coefficients of i.i.d. random variables \cite{Brillinger75,BriRos67}.
The Gaussian integrals are easily done.  Still letting
{\bf 1} represent the $2 \times 2$ unit matrix one finds that 
\beq
p_n(\alpha) \simeq  C(\alpha)p_n^{(0)}(\alpha), \qquad
C(\alpha)= \prod_{q=1}^\infty\, \Lambda_q^{-1}(\alpha), \qquad 
\alpha>1,
\label{respnalpha}
\eeq
where we abbreviated for all $\alpha \geq 1$ 
\bea 
\Lambda_q(\alpha) &=& \det(\bse+\bsV_q\bsE)   \nonumber\\[1mm] 
&=& 1 + 3A_q -2B_q -\tfrac{1}{2}B_q^2 \nonumber\\[1mm]
&=& 1 - (3-\alpha)q^{-2} + \alpha^2 q^{-4}.
\label{defmuq}
\eea
Equations (\ref{respnalpha}) and (\ref{defmuq}) 
are the main result of this section.
It is easily verified that $\Lambda_q(\alpha) > 0$ for all
$\alpha\geq 1$ and $q=1,2,\ldots$
with the only exception that $\Lambda_1(1)=0$.
This is the reason why the substitution 
that led to (\ref{pnaspt})  
must be restricted to $\alpha>1$;
for the Crofton point $\alpha=1$ the resulting
Gaussian integration would diverge.
Hence the method of this section leaves the Crofton problem unsolved.
In the following section we will develop a modified
method suitable for the Crofton cell.

%%%%%%%%%%%%%%%%%%%%%%%%%%%%%%%%%%%%%%%%%%%%%%%%%%%%%%%%%%%%%%%%%%%%%%%%%%%%
%%%%%%%%%%%%%%%%%%%%%%%%%%%%%%%%%%%%%%%%%%%%%%%%%%%%%%%%%%%%%%%%%%%%%%%%%%%%

\section{The Crofton cell: $\alpha=1$}
\label{secCrofton}

\subsection{Centering condition}
\label{secstart}
%%% [*CR 1121]

For $\alpha=1$ the expansion of
$\la\Theta\ee^{-{\mathbb V}(\alpha)}\ra$ around a regular $n$-gon 
leads to a diverging integral and hence cannot be used for finding $p_n(1)$.
This divergence, due to the vanishing of $\Lambda_q(1)$ with $q=1$,
may be tracked down to the origin being located 
with the same probability in any point of the cell;
hence the problem has a `zero mode' due to translational invariance.
Indeed, for $q=\pm 1$ elastic deformations of a circle are actually
translations.

This analysis also points the way out of the difficulty.
In the integrals one should count a Crofton cell  
only when, under a suitable definition, it is `centered' around the origin, 
and then give it a weight proportional to its area 
in order to simultaneously account for
all noncentered cells that may be obtained from it by translation. 
The `centering condition' that we adopt is that the $q=\pm 1$ 
Fourier components of $R_m$ vanish, that is,
\beq
\hat{R}_1 \equiv \frac{1}{n} \summ \ee^{2\pi {\rm i}m/n} R_m = 0.
\label{centeringcond}
\eeq
This is a constraint only on the vector lengths $R_m$, irrespective
of the angles $\Phi_m$.

The following remark establishes one more link with the mathematical
literature. There, a function ${\cal A}_2(\bRn;\mathbf{r})$ 
appears defined as 
${\cal A}_\alpha(\bRn)$ with $\alpha=2$
in (\ref{defcalAalpha}) except that the support
function $h(\Phi)$ with respect to the origin $\bf O$ 
is replaced by the support function $h(\Phi;\mathbf{r})$
with respect to $\mathbf{r}$. The function ${\cal A}_2(\bRn;\mathbf{r})$ 
is commonly used to calculate the
probability that a convex set containing the origin translated by
$-\mathbf{r}$ is included in the typical Poisson-Voronoi cell. As a function
of $\mathbf{r}$, ${\cal A}_2(\bRn;\mathbf{r})$ is a convex function such that
its gradient $\nabla_{\!\mathbf{r}}\,{\cal A}_{2}(\bRn;\mathbf{r})$ is,
to, up to a multiplicative constant, equal to
$\big( \int_0^{2\pi}h(\Phi;\mathbf{r})\cos\Phi d\Phi,\,
       \int_0^{2\pi}h(\Phi;\mathbf{r})\sin\Phi d\Phi \big).$   
The sum appearing in (\ref{centeringcond}) is the discrete
version of this gradient. 
Hence (\ref{centeringcond}) amounts to adopting as the
`center' of the cell the point for which the fundamental domain has minimum
area; we remark that this point is given by
$\big( {\pi}^{-1}\int_0^{2\pi}h(\Phi)\cos\Phi\, d\Phi,\,
       {\pi}^{-1}\int_0^{2\pi}h(\Phi)\sin\Phi\, d\Phi \big)$.  

In the presence of constraint (\ref{centeringcond})
the necessary mathematics requires a few preliminaries.
Let $\bt=(t_x,t_y)$ be a translation applied to a Crofton cell.
Under $\bt$  the vertices $\bS_m$ are translated by just that amount.
However, the $\bR_m$, being projections of the origin
onto lines that under the translation stay parallel to themselves,
transform differently.
If $\bt$ keeps the origin inside the cell, the $\bR_m$
transform into $\bR_{m,\mathbf{t}}=(R_{m,\mathbf{t}},\Phi_{m,\mathbf{t}})$
given by
\bea 
R_{m,\mathbf{t}} &=& R_m + t_x\cos\Phi_m + t_y\sin\Phi_m, \nonumber\\
\Phi_{m,\mathbf{t}} &=& \Phi_m.
\label{translation}
\eea
We will write $\bRnbt=\{\bR_{1,\mathbf{t}},\ldots,\bR_{n,\mathbf{t}}\}$.
It will be convenient to extend the definition (\ref{translation}) 
of $\bRnbt$ to translations $\bt$ that take the origin out of the cell. 
In that case, however, $\bRnbt$ is `nonphysical' 
(it does not represent a Crofton cell any more;
one or more of its $R_{m,\mathbf{t}}$ are negative)
and, by virtue of (\ref{defchi0G}),
the indicator $\chi(\bRnbt)$ then vanishes.

We now show that if $\bRn$ defines a Crofton cell, 
then there exists a unique `centering translation' $\bt^*(\bRn)$, 
that is, one for which (\ref{centeringcond}) is satisfied: 
\beq
\hat{R}_{1,\mathbf{t}^*} = 0.
\label{defTstar}
\eeq
The proof goes by explicit construction:
%%% [*CR 1111; 1202]
using (\ref{centeringcond}) and (\ref{translation}) in (\ref{defTstar})
leads to a linear system of equations for $\bt^*$ whose solution is
\beq
\bt^*(\bRn) = 
-\tfrac{1}{2}\sqrt{2} n\,{\bsphi}^{-1} 
\binom{ \hat{R}_1^{\rm c} }{ \hat{R}_1^{\rm s} },
\label{solTstar}
\eeq
where for any complex variable $z$ we set 
$z = ( z^{\rm c} + {\rm i}z^{\rm s} )/\sqrt{2}$
and where $\bsphi$ is the matrix of elements
\bea 
&& \varphi_{11}=\sum_m\cos(2\pi m/n)\cos\Phi_m, \qquad
\varphi_{12}=\sum_m\cos(2\pi m/n)\sin\Phi_m, \nonumber\\
&& \varphi_{21}=\sum_m\sin(2\pi m/n)\cos\Phi_m, \qquad
\varphi_{22}=\sum_m\sin(2\pi m/n)\sin\Phi_m\,. \phantom{xxx}
\label{defphi}
\eea
Two useful relations,
\beq
\bt^*(\bRnbt) = \bt^*( \bRn ) - \bt,
\label{relTstar}
\eeq
and
\beq
\int_{{\mathbb R}^2}\! \dd\bt\,
\delta\big( \bt-\bt^*(\bRn) \big) \chi(\bRnbt) = 1.
\label{deltainsert}
\eeq
%%% [*CR 1123]
will serve below.

%%%%%%%%%%%%%%%%%%%%%%%%%%%%%%%%%%%%%%%%%%%%%%%%%%%%%%%%%%%%%%%%%%%%%%%%%%%%%%

\subsection{Modified starting point for $\alpha=1$}
\label{secnewstart}

After these preliminaries we return to the general expression (\ref{defpnbis})
for the sidedness probability $p_n(\alpha)$. 
For the case of the Crofton cell we insert
(\ref{deltainsert}) and rewrite $p_n(1)$ as
\bea 
p_n(1) &=& \frac{\lambda^n}{n!}\int_0^{2\pi}\! \dd\Phi_1\ldots\dd\Phi_n
\int_{{\mathbb R}^2}\! \dd\bt \int_0^\infty \dd R_1\ldots\dd R_n\,
\nonumber\\[2mm]
&& \times\,
\delta \big( \bt-\bt^*(\bRn) \big)\,
\chi(\bRn)\,\chi(\bRnbt)\,\ee^{-{\cal P}(\bRn)},
\label{defpn1}
\eea
where we used that ${\cal A}_1={\cal P}$ and have indicated explicitly the
dependence on $\bRn$ of all quantities involved.

Under the $\bt$ integral in (\ref{defpn1}) 
we now transform from the $R_m$ to new radial
variables of integration $R'_m \equiv R_{m,\mathbf{t}}$, 
with the $R_{m,\mathbf{t}}$ given by (\ref{translation}).
We may express $\bRn$ in terms of $\bRprimen$ by means of $\bRn=\bRprimenmbt$.
The Jacobian of this transformation
is unity but the limits of integration should be treated with some care. 
Equation (\ref{translation}) implies for $R'_m$ the domain of 
integration $t_x\cos\Phi_m + t_y\sin\Phi_m <R'_m<\infty$.
If the lower limit of integration is negative, we may replace it with zero
because $\chi(\bRnbt)=0$ on the interval discarded; 
and if the lower integration limit is positive, 
we may also replace it with zero, because then $\chi(\bRn)=0$ on the interval
added to the domain of integration.
Hence, still using (\ref{relTstar}), we get
\bea 
p_n(1) &=& \frac{\lambda^n}{n!}\int_0^{2\pi}\! \dd\Phi_1\ldots\dd\Phi_n
\int_{{\mathbb R}^2}\! \dd\bt \int_0^\infty 
\dd R_1^\prime\ldots\dd R_n^\prime\,
\nonumber\\[2mm]
&& \times\,
\delta \big( \bt^*(\bRprimen) \big)\,
\chi(\bRprimen)\,\chi(\bRprimenmbt)\,
\ee^{-{\cal P}(\bRprimenmbt)}.
\label{defpn1bis}
\eea
Since the cell perimeter is invariant under translation, we have 
${\cal P}(\bRprimenmbt)={\cal P}(\bRprimen)$.
The only $\bt$ dependence left in the integrand of (\ref{defpn1bis})
is then the one in $\chi(\bRprimenmbt)$. The integral on $\bt$ of this
quantity has a nonzero contribution only in a domain 
of the same size and shape as the cell itself
and therefore produces the cell area $A(\bRprimen)$. 
Hence, still suppressing the primes, we get from (\ref{defpn1bis})
\bea
p_n(1) &=& \frac{\lambda^n}{n!}\int_0^{2\pi}\! \dd\Phi_1\ldots\dd\Phi_n
\int_0^\infty \dd R_1\ldots\dd R_n
\nonumber\\[2mm]
&& \times\,
\delta \big( \bt^*(\bRn) \big)\,
\chi(\bRn)\,A(\bRn)\,\ee^{-{\cal P}(\bRn)},
\label{defpn1ter}
\eea
Equation (\ref{defpn1ter}) constitutes a modified starting point 
for the calculation of the sidedness probability $p_n(1)$
of the Crofton cell. It differs from the original expression (\ref{defpnbis})
by the insertion of a delta function and of the area factor $A(\bRn)$. 

%%%%%%%%%%%%%%%%%%%%%%%%%%%%%%%%%%%%%%%%%%%%%%%%%%%%%%%%%%%%%%%%%%%%%%%%%%%%
%%%%%%%%%%%%%%%%%%%%%%%%%%%%%%%%%%%%%%%%%%%%%%%%%%%%%%%%%%%%%%%%%%%%%%%%%%%%

\subsection{Sidedness probability $p_n(1)$ of the Crofton cell}
\label{secsidedness}

Having rewritten the definition of the sidedness probability
$p_n(1)$ as (\ref{defpn1ter}),
we are now in a position to start its explicit evaluation. 
The cell area $A$ is given by
\beq
A = R_{\rm av}^2 A_{\rm ang}, \qquad
A_{\rm ang} = \tfrac{1}{2} \summ \rho_m^2 (\tan\gamma_m + \tan\beta_{m+1}), 
\label{defAang}
\eeq 
where $A_{\rm ang}$ is an expression only in terms of the angular variables.
We may similarly isolate the radial part
of the centering translation $\bt^\star$ found in 
(\ref{solTstar}) by writing
\beq
\bt^\star = R_{\rm av}^2 \bt_{\rm ang}^\star,
\label{defTstarang}
\eeq
where $\bt_{\rm ang}^\star$ depends only on the angles.

When we substitute (\ref{defAang}) and (\ref{defTstarang}) in 
(\ref{defpn1ter}), the factors $R_{\rm av}$ stemming from 
$\delta(\bt^\star)$ and from the area $A$ cancel.
We set ${\cal P} = 2\pi R_{\rm av}(1+n^{-1}V)$, which is the special case
$\alpha=1$ of (\ref{exprcalAalpha}); here $2\pi R_{\rm av}$ is the
perimeter of a circular cell of radius $R_{\rm av}$.
We may in the same way as before pass to the variables $\{\xi_m,\eta_m\}$
and a single length scale $R_{\rm av}$.
Carrying out the $R_{\rm av}$ integration we get, setting $\lambda=1$ as
before, 
\bea
p_n(1) &=& \frac{1}{n!} \int_0^{2\pi}\! \dd\Phi_1\ldots\dd\Phi_n
\int_0^\infty \dd R_1\ldots\dd R_n\,
\delta ( \bt^*_{\rm ang} )\,A_{\rm ang}\,
\chi\,\ee^{-{\cal P}}
\nonumber\\
&=&
\frac{1}{n} \int_{\xi,\eta}
G'(\xi,\eta;\beta_*)^{-1}
\Big[\prod_{m=1}^n \rho_m^\alpha T_m^{\phantom{2}}\xi_m^{-1} \Big] 
\,\Theta\,\int_0^\infty \!\dd R_{\rm av}\, R_{\rm av}^{n -1}\,\,
\ee^{-{\cal P}}
\nonumber\\
&=&
\frac{(n-1)!}{n(2\pi)^n} \int_{\xi,\eta} 
\Theta\,\delta(\bt^\star_{\rm ang})\,A_{\rm ang}\,\ee^{-{\mathbb V}(1)},
\label{pnafterint1}
\eea
which is an alternative for (\ref{pnafterint}) when $\alpha=1$.
We may rewrite (\ref{pnafterint1}) as
\beq
p_n(1) = p_n^{(0)}(1)\, \la 
\Theta\,\delta(\bt^*_{\rm ang})\,A_{\rm ang}\,\ee^{-{\mathbb V}(1)} \ra,
\label{pn1exact}
\eeq
where $p_n^{(0)}(1)$ and
${\mathbb V}(1)$ are given by the same equations as before, namely
(\ref{calcpn0}) and (\ref{defmathbbV}) with $\alpha=1$,
and the angular brackets are defined by (\ref{deflara}).
When $\alpha=1$,
expression (\ref{pn1exact}) is an alternative to (\ref{pnfactors}).
Both are exact representations of the initial
expression for $p_n(1)$, but
the (\ref{pn1exact}) is required if the
large-$n$ expansion is to succeed. 
That will be the subject of section \ref{secexpansion}.

%%%%%%%%%%%%%%%%%%%%%%%%%%%%%%%%%%%%%%%%%%%%%%%%%%%%%%%%%%%%%%%%%%%%%%%%%%%%%

\subsection{Large-$n$ expansion of $p_n(1)$}
\label{secexpansion1}

%%%%%%%%%%%%%%%%%%%%%%%%%%%%%%%%%%%%%%%%%%%%%%%%%%%%%%%%%%%%%%%%%%%%%%%%%%%%%

\subsubsection{Preliminaries}
\label{secpreliminary}

We will perform the large-$n$ expansion of the average appearing in
(\ref{pn1exact}).
The expansion of ${\mathbb V}(1)$ is not different from the general case and
the result follows directly from 
(\ref{expmathbbV})-(\ref{defAqBq}) by setting $\alpha=1$.
As before, the indicator $\Theta$ will give corrections that are
exponentially small with $n$; we may therefore replace it with
unity. 
The insertion $A_{\rm ang}$ in (\ref{pn1exact}), 
given explicitly in (\ref{defAang}), is expanded as
\bea
A_{\rm ang} &=& \tfrac{1}{2}\summ
\big[ \gamma_m + \beta_{m+1} + {\cal O}(\gamma_m^3,\beta_{m+1}^3) \big]
\nonumber\\
&=& \pi + {\cal O}(n^{-\half}),
\label{expAang}
\eea
which, to leading order, is the area of a disk of unit radius.
We now investigate the factor $\delta(\bt^*_{\rm ang})$ for large $n$.
In view of (\ref{defTstarang}) and (\ref{solTstar}) this requires
the asymptotic evaluation of the matrix $\bsphi$ defined in (\ref{defphi}).
In the large-$n$ limit the angle differences                                   
$\xi_\ell=\Phi_\ell-\Phi_{\ell-1}$
are independent random variables of average $2\pi/n$.
Hence after setting $\Phi_0 = 0$ and summing $\xi_\ell$
on $\ell$ from $1$ to $m$ we have $\Phi_m=2\pi m/n + {\cal O}(n^{-\half})$.
%(where the correction term is stochastic). 
By doing the sums in (\ref{defphi}) one then finds that $\bsphi$ is equal
to $\tfrac{1}{2} n$ times the $2 \times 2$ unit matrix, 
up to corrections that vanish for
$n\to\infty$. Setting  $\hat{R}_1 = R_{\rm av} \hat{\rho}_1$ we have
\beq
\delta(\bt^*_{\rm ang}) \simeq 
\tfrac{1}{2} \delta(\hat{\rho}_1^{\rm c})\delta(\hat{\rho}_1^{\rm s}),
\label{deltatang}
\eeq
where the symbol $\simeq$ indicates validity for $n\to\infty$.
In that limit we can express $\hat{\rho}_1$ 
in terms of the $\hat{X}_q$ and $\hat{Y}_q$ by means of the relation
\cite{Hilhorst05b}
\beq
\hat{\rho}_1 \simeq n^\half (\hat{X}_1 - \hat{Y}_1).
\label{relrhoXY}
\eeq
Using (\ref{expAang})-(\ref{relrhoXY}) in (\ref{pn1exact}) we find that
\beq
p_n(1) \simeq  p_n^{(0)}(1) 
\,\times\,\tfrac{1}{2}\pi n\,
\la 
\delta(\hat{X}^{\rm c}_1-\hat{Y}^{\rm c}_1)
\delta(\hat{X}^{\rm s}_1-\hat{Y}^{\rm s}_1)\,\ee^{-{\mathbb V}_1(1)} 
\ra,
\label{finalX1Y1}
\eeq
where ${\mathbb V}_1(1)$ is given by (\ref{resV1}) with $\alpha=1$ and
$\la\ldots\ra$ is the average with respect to the Gaussian weight 
(\ref{Gaussianweight}). 
\vspace{3mm}

%%%%%%%%%%%%%%%%%%%%%%%%%%%%%%%%%%%%%%%%%%%%%%%%%%%%%%%%%%%%%%%%%%%%%%%%%%%%%

\subsubsection{The Gaussian integrations}
\label{secfinalintegrations}

In (\ref{finalX1Y1}) the integrations on the Fourier variables 
$\hat{X}_q$ and $\hat{Y}_q$ with $q \neq \pm 1$ may be carried out in the
same way as before and lead to
\bea
p_n(1) &\simeq& p_n^{(0)}(1) \times \tfrac{1}{2}\pi n\, 
\prod_{q=2}^\infty (1-q^{-2})^{-2} \nonumber\\[2mm]
&& \times
\la 
\delta(\hat{X}^{\rm c}_1-\hat{Y}^{\rm c}_1)
\delta(\hat{X}^{\rm s}_1-\hat{Y}^{\rm s}_1)\,
\ee^{ -\hat{X}_1\hat{X}_{-1} + \hat{Y}_1\hat{Y}_{-1} } 
\ra,
\label{intermediate}
\eea
where the terms in the exponential represent the $q=\pm 1$ contribution to
${\mathbb V}_1(1)$.
The average $\la\ldots\ra$ in (\ref{intermediate}) reads explicitly 
\beq
\la\ldots\ra = \frac{1}{2\pi^2}\int\dd\hat{X}_1 \dd\hat{X}_{-1}
\dd\hat{Y}_1 \dd\hat{Y}_{-1}\,\ldots\,
\ee^{ -2\hat{X}_1\hat{X}_{-1} - \hat{Y}_1\hat{Y}_{-1} },
\label{Gweight1}
\eeq
as follows from (\ref{Gaussianweight}) 
when restricted to its $q=\pm 1$ Fourier components.
When (\ref{Gweight1}) and (\ref{intermediate}) are combined,
the terms $\hat{Y}_{1}\hat{Y}_{-1}$ in the exponents cancel 
and the average in (\ref{intermediate}) turns out to be equal to $1/(3\pi)$.
Without the delta function insertions in the angular brackets,
these final integrations on $\hat{Y}_{\pm 1}$ would have led to a
divergence. 
The fact that they now remain finite confirms the validity of our approach.

After combining everything and still using that
$\prod_{q=2}^\infty (1-q^{-2}) = \tfrac{1}{2}$,
we obtain the main result of this section: the sidedness probability $p_n(1)$
of the Crofton cell is given by
\beq
p_n(1) \simeq
\tfrac{2}{3} n \,p_n^{(0)}(1), \qquad n\to\infty,
\label{resfinpn1}
\eeq
with $p_n^{(0)}(1)$ given by (\ref{resultpn0}).
This is what was announced in 
section \ref{secresults} of the introduction.

%%%%%%%%%%%%%%%%%%%%%%%%%%%%%%%%%%%%%%%%%%%%%%%%%%%%%%%%%%%%%%%%%%%%%%%%%%%

\section{The typical cell in a Poisson line tessellation}
\label{sectypical}
%%% [* CR1041]

In this section we continue our study of the case $\alpha=1$, that is, the
Poisson line tessellation. But whereas in section \ref{secCrofton}
we studied the Crofton cell (or zero-cell), we will now consider the typical
cell.
The preceding results give rise to a corollary concerning the typical cell
that can be derived with little effort.
Let $p_n^{\rm typ}(1)$ denote the sidedness probability
of the typical cell in the Poisson line tessellation.
As was shown in detail by Calka \cite{Calka03a}, 
the expression for $p_n^{\rm typ}(1)$ differs
from equation (\ref{defpn1}) for $p_n(1)$ by the
insertion in the integrand of the latter of an extra factor 
$\la{A}\ra^{\rm typ}/A(\bRn)$, 
where $\la{A}\ra^{\rm typ}=1/\pi\lambda^2$ 
is the average cell area \cite{Goudsmit45} in the Poisson line tessellation.
%%% [* CR 1036]

The extra $1/A(\bRn)$ cancels the $A(\bRn)$ present in (\ref{defpn1ter})
and hence 
\beq
p_n^{\rm typ}(1) = 
\frac{\lambda^{n-2}}{\pi n!}\int_0^{2\pi}\! \dd\Phi_1\ldots\dd\Phi_n
\int_0^\infty \dd R_1\ldots\dd R_n
\,\delta \big( \bt^*(\bRn) \big)\,
\chi(\bRn)\,\ee^{-{\cal P}(\bRn)}.
\label{defpn1typ}
\eeq
In the limit $n\to\infty$ this is easily evaluated by the methods of sections
\ref{secsidedness} and \ref{secexpansion1}.
The integral on $\Rav$ now has an extra factor $\Rav^{-2}$ in its integrand
and as a consequence we get instead of (\ref{pn1exact})
the expression
\beq
p_n^{\rm typ}(1)=\frac{4\pi}{(n-1)(n-2)}\, 
\la \Theta\,\delta(\bt^*_{\rm ang})\,( 1+n^{-1}V )^{-2}\,
\ee^{-{\mathbb V}(1)} \ra,
\label{pn1typexact}
\eeq
where we have again set $\lambda=1$.
Equation (\ref{pn1typexact}) is still a fully exact expression 
for the sidedness of the typical cell. 
>From here on, again, we have to resort to a large-$n$ expansion. 
Since 
$( 1+n^{-1}V )^{-2}=1+{\cal O}(n^{-1})$,
the average in (\ref{pn1typexact}) is to leading order in $n$  
identical to the one in
(\ref{pn1exact}) except for the absence of the insertion $A_{\rm ang}$, 
which to leading order is equal to $\pi$.
Upon combining these considerations we get from (\ref{pn1typexact}) 
\beq
p_n^{\rm typ}(1) \simeq \tfrac{8}{3}n^{-1}p_n^{(0)}(1), \qquad n\to\infty.
\label{resfinpn1typ}
\eeq
This is our final result, announced in section \ref{secresults},
for the asymptotic sidedness probability of the typical cell.
Comparison of (\ref{resfinpn1}) and (\ref{resfinpn1typ}) shows that
the zero-cell and the typical cell sidedness probabilities have the ratio
\beq
\frac{p_n(1)}{p_n^{\rm typ}(1)} \,\simeq\, \tfrac{1}{4}n^2 \,=\, 
\frac{\pi\Rc^2}{\la{A}\ra^{\rm typ}}, \qquad n\to\infty,
\label{ratioptypp}
\eeq
where $\pi\Rc^2 = n^2/4\pi$ [see (\ref{defRcsigmac})]
is the average area of an $n$-sided Crofton cell  
and $\la{A}\ra^{\rm typ}=1/\pi$ is the average typical-cell area.
Equation (\ref{ratioptypp}) is heuristically obvious.

%%%%%%%%%%%%%%%%%%%%%%%%%%%%%%%%%%%%%%%%%%%%%%%%%%%%%%%%%%%%%%%%%%%%%%%%%%%

\section{Properties beyond $p_n$}
\label{secother}

%%%%%%%%%%%%%%%%%%%%%%%%%%%%%%%%%%%%%%%%%%%%%%%%%%%%%%%%%%%%%%%%%%%%%%%%%%%%%

\subsection{General}
\label{secgeneralproperties}

In this section we return to a general value of the parameter $\alpha$.
The analysis of sections \ref{secfamily} and \ref{secCrofton}
focused on finding the sidedness probability $p_n$.
It started from a weight functional for the zero-cell expressed initially [in
(\ref{defpn})] as $\chi\exp(-{\cal A}_\alpha)$.
In the course of the analysis we were led to transform 
this functional into $\exp(-{\mathbb V}_1)$ where ${\mathbb V}_1$ is 
quadratic in the Fourier variables $\hat{X}_q$ and $\hat{Y}_q$, which in turn 
are Gaussian distributed according to (\ref{Gaussianweight}).
The transformation is valid in the limit $n\to\infty$
and the resulting weight $\exp(-{\mathbb V}_1)$ may be used to
calculate arbitrary averages of functionals of the cell perimeter,
provided that sums and products on the wavenumber $q$ converge
sufficiently rapidly. 
We will limit ourselves below to discussing only one of these cell averages.

%%%%%%%%%%%%%%%%%%%%%%%%%%%%%%%%%%%%%%%%%%%%%%%%%%%%%%%%%%%%%%%%%%%%%%%%%%%%%

\subsection{Fluctuations of  $R_m$ around $\Rav$}
\label{secfluctsRm}
%%% [*CR 1013]

Important information on the cell shape is contained in the 
root mean square deviation $\sigma_R$
of the vectors $R_m$ from their average $\Rav$.
Straightforward calculation \cite{Hilhorst05b} yields  
for this quantity the expression
\bea
\sigma_R^2 &\equiv&
n^{-1}\sum_{m=1}^n\la(R_m-\Rav)^2 \ra 
\nonumber\\[2mm]
&\simeq& 
2n^{-1}R_{\rm c}^2\sum_{q=1}^\infty q^{-4}\la|\hat{X}_q-\hat{Y}_q|^2\ra,
\qquad n\to\infty.
\label{defsigmaR}
\eea
It only requires a Gaussian integration to find that for $n\to\infty$
\beq
\la|\hat{X}_q-\hat{Y}_q|^2\ra \simeq \tfrac{3}{2}\Lambda_q^{-1}(\alpha),
\qquad q= 2, 3,\ldots, \quad \alpha \geq 1.
\label{exprcorr}
\eeq
A special case is
\beq
\la|\hat{X}_1-\hat{Y}_1|^2\ra \simeq 
\left\{
\begin{array}{ll}
\tfrac{3}{2}\Lambda_1^{-1}(\alpha), & \alpha>1,\\[2mm]
0, & \alpha=1,
\end{array}
\right.
\label{exprcorr1}
\eeq
where the discontinuity at $\alpha=1$ is due to the centering condition
that we applied in that case.
Substitution in (\ref{defsigmaR}) of (\ref{exprcorr}) and (\ref{exprcorr1}), 
as well as of the explicit expression (\ref{defRcsigmac}) for $R_{\rm c}$, 
shows that 
\beq
\sigma_R \simeq
\left\{
\begin{array}{ll}
c(\alpha)\,  n^{^{\frac{1}{\alpha}-\frac{1}{2}}}, & \alpha>1,\\[2mm]
\bar{c}(1)\, n^{^\half}, & \alpha=1,
\end{array}
\right.
\label{resultsigmaR}
\eeq
where the coefficients have the properties
\bea
c(\alpha) &\simeq& \frac{1}{2\pi(\alpha-1)^\half}\,,  
\qquad \alpha\to 1, \nonumber\\[2mm]
\bar{c}(1) &=& \frac{3}{2\pi^2}\sum_{q=2}^\infty\frac{1}{(q^2-1)^2}
=\tfrac{1}{16}\big( 1\,-\,\tfrac{33}{4\pi^2} \big) = 0.010256...
\label{propc}
\eea
The divergence of $c(\alpha)$ as $\alpha\to 1$ 
is due to the divergence of the
$q=1$ term in the sum in (\ref{defsigmaR}); it demonstrates once more
the tendency of the center of the cell to become delocalized from the origin
in that limit. For $\alpha=1$ we obtain discontinuously
the value of $\bar{c}(1)$ due to our centering of the cell. 

%%%%%%%%%%%%%%%%%%%%%%%%%%%%%%%%%%%%%%%%%%%%%%%%%%%%%%%%%%%%%%%%%%%%%%%%%%%%%%%

\section{Conclusion and final remarks}
\label{secconclusion}

We have determined the asymptotic large-$n$ expansion
of the sidedness probability $p_n(\alpha)$ of the zero-cell
in a family of tessellations dependent on a parameter $\alpha$.
Special cases are the typical Poisson-Voronoi cell ($\alpha=2$),
and the Crofton cell ($\alpha=1$). The latter turned out to be a singular
point in the parameter range and required an extension
of previously known methods.

We end with a few remarks
about further research to which this work opens the door.
The first remark concerns the parameter range $0<\alpha<1$.
In this range the density $\rho(\bR)$ and the
probability distribution $P(\bR)$ defined in (\ref{defP}) are still
integrable in the origin. 
However, $\Lambda_1(\alpha)$ is negative,
which indicates a tendency for the origin to be away from the cell
center defined by equation (\ref{centeringcond}).
Exploring the range $0<\alpha<1$ further would definitely be of interest;
it is no longer certain that in that case, too, the zero-cell will tend to a
circle in the large-$n$ limit.

Secondly, the analytic method of this work
points the way to a new Monte Carlo simulation algorithm for 
the Crofton cell, similar to the one that was used
\cite{Hilhorst07} successfully for the Voronoi cell,
and capable of producing accurate $p_n$ values for all finite $n$.

Thirdly, one may consider sidedness {\it correlations\,} between
distinct cells, whether neighboring or separated by a larger distance.
In the Voronoi case ($\alpha=2)$, the correlation between 
adjacent cells is commonly denoted $m_n$ and defined as the average sidedness
of a cell given that its neighbor is $n$-sided.
To our knowledge, no one has yet attempted to determine
the analog of this quantity for the Poisson line tessellation.

We will leave these matters for the future.

%%%%%%%%%%%%%%%%%%%%%%%%%%%%%%%%%%%%%%%%%%%%%%%%%%%%%%%%%%%%%%%%%%%%%%%%%%%%
%%%%%%%%%%%%%%%%%%%%%%%%%%%%%%%%%%%%%%%%%%%%%%%%%%%%%%%%%%%%%%%%%%%%%%%%%%%%
%%%%%%%%%%%%%%%%%%%%%%%%%%%%%%%%%%%%%%%%%%%%%%%%%%%%%%%%%%%%%%%%%%%%%%%%%%%%

\appendix

%%%%%%%%%%%%%%%%%%%%%%%%%%%%%%%%%%%%%%%%%%%%%%%%%%%%%%%%%%%%%%%%%%%%%%%%%%%%%%
\section*{Acknowledgments}
The second author wishes to acknowledge the support of the French ANR Project
``mipomodim'' No. ANR-05-BLAN-0017.  

%%%%%%%%%%%%%%%%%%%%%%%%%%%%%%%%%%%%%%%%%%%%%%%%%%%%%%%%%%%%%%%%%%%%%%%%%%%%%%

\end{document}